# *Myxococcus xanthus* gliding motors are elastically coupled to the substrate as predicted by the focal adhesion model of gliding motility

Rajesh Balagam[1], Douglas B. Litwin[2], Fabian Czerwinski[3], Mingzhai Sun[3], Heidi B. Kaplan[2], Joshua W. Shaevitz*[3], Oleg A. Igoshin*[1]

[1]Department of Bioengineering, Rice University, TX, USA.

[2]Department of Microbiology and Molecular Genetics, University of Texas – Houston Medical School, Houston, TX, USA.

[3]Department of Physics and the Lewis-Sigler Institute of Integrative Genomics, Princeton University, Princeton, NJ, USA.

*Co-corresponding authors, E-mails: Shaevitz@princeton.edu, igoshin@rice.edu

**Short title:** Discriminating between *M. xanthus* motility models

## Abstract

*Myxococcus xanthus* is a model organism for studying bacterial social behaviors due to its ability to form complex multi-cellular structures. Knowledge of *M. xanthus* surface gliding motility and the mechanisms that coordinated it are critically important to our understanding of collective cell behaviors. Although the mechanism of gliding motility is still under investigation, recent experiments suggest that there are two possible mechanisms underlying force production for cell motility: the focal adhesion mechanism and the helical rotor mechanism which differ in the biophysics of the cell – substrate interactions. Whereas the focal adhesion model predicts an elastic coupling, the helical rotor model predicts a viscous coupling. Using a combination of computational modeling, imaging, and force microscopy, we find evidence for elastic coupling in support of the focal adhesion model. Using a biophysical model of the *M. xanthus* cell, we investigated how the mechanical interactions between cells are affected by interactions with the substrate. Comparison of modeling results with experimental data for cell-cell collision events pointed to a strong, elastic attachment between the cell and substrate. These results are robust to variations in the mechanical and geometrical parameters of the model. We then directly measured the motor-substrate coupling by monitoring the motion of optically trapped beads and find that motor velocity decreases exponentially with opposing load. At high loads, motor velocity approaches zero velocity asymptotically and motors remain bound to beads indicating a strong, elastic attachment.

## Significance (120 words limit)

Studies of collective bacterial motility on solid surfaces are essential for understanding self-organization of biofilms. The Gram-negative bacterium *Myxococcus xanthus* has long been used as a model organism for studying surface motility but its mechanisms of gliding motility are still under investigation. Recent experiments point to two potential mechanisms that differ qualitatively in the details of cell-substrate interactions. To investigate the biophysical nature of this interaction, we developed a synergistically multidisciplinary approach combining computational modeling, time-lapse microscopy, and biophysical optical trap experiments. The results conclusively showed strong adhesive attachments between cell and substrate, providing support for an elastic rather than viscous coupling between cell and substrate similar to phenomena observed in focal adhesions from eukaryotic cells.

## Introduction

*Myxococcus xanthus* is a predatory soil bacterium that has been widely used as a model organism for studies of bacterial social behaviors (1). Under different environmental conditions *M. xanthus* cells display a range of complex multi-cellular behaviors, including groups of cells moving together (often referred to as swarms), periodic bands of high cell density travelling waves (termed ripples), and aggregates of more than $10^5$ cells containing environmentally-resistant myxospores (termed fruiting bodies) (2). Formation of these complex self-organized patterns requires coordination and collective motility among the cells. The biophysical mechanisms underlying the cell motility and intercellular interactions that generate these collective behaviors are still not completely understood.

*M. xanthus* cell movement is limited to translocation on solid surfaces using two different flagella-independent motility systems (3). Gliding motility, previously termed adventurous (A) motility, is defined as active surface translocation along the long cell axis without the aid of flagella or pili and is responsible for individual cell movement. Twitching motility, previously termed social (S) motility, appears similar to gliding motility, but is limited to cells within at least a cell length of another cell and is known to be powered by type IV pili extension and retraction (4). The biophysical mechanism of gliding motility in *M. xanthus* and other bacteria is the subject of active research.

Earlier studies on the mechanism of gliding motility hypothesized that the exopolysaccharide (EPS) slime secretion at the cell's lagging pole and the expansion of slime due to hydration was responsible for the motility (5-7). However, subsequent experimental studies (8, 9), indicated that force generation in gliding motility is likely to be distributed along the cell length.

Recently, an alternative view of the gliding motility mechanism has emerged. Using fluorescently tagged proteins recent experiments identified a few components of the machinery responsible for the distributed force-generation: gliding motility regulatory protein (AglZ) (10) and motor proteins (AglRQS) (11). These studies showed clustering of these proteins at regular intervals along the cell length. These clusters appear to form at the cell's leading pole and disperse at the lagging pole, while remaining stationary with respect to the substrate during cell movement. Further, depolymerization of the cell cytoskeleton elements (MreB) dispersed these clusters and inhibited the gliding motility (11). Based on the above observations, a focal adhesion mechanism (FAM) of gliding motility was proposed (10, 11) (Fig. 1A). The mechanism hypothesizes that intracellular motor proteins moving on helical cytoskeletal filaments are somehow connected to the focal adhesion complexes attached to a substrate. The cell movement is therefore generated by motors pushing against these focal adhesion complexes. However, it is not clear which molecules adhere cells to the substrate and how their connection to the motor complex is able to move through the peptidoglycan of cell wall.

Another study observed that AgmU, a gliding motility protein, is part of a multi-protein complex that spans cell's inner membrane and periplasm (12). Additionally, it was found that AgmU decorates a looped continuous helix that rotates as the cell moves forward (13). It was shown that rotation of the helix stopped when MreB cytoskeletal filaments were depolymerized (13). The authors also observed that a periodic distortion of cell wall that is consistent with periodicity of the MreB helices. Based on these observations, a helical rotor mechanism (HRM) (5, 13) (Figure 1B) of gliding motility was proposed. In this mechanism motor proteins (AglR) (14) distort cell surface by interacting with the gliding motility proteins (AglZ, AgmU) in protein clusters and create drag forces between cell surface and substrate. These drag forces propel the cell forward.

Even though these studies provide ample evidence for both FAM and HRM mechanisms of gliding motility, neither mechanism has been conclusively proven or eliminated. We note that the major biophysical distinction between the mechanisms is in the nature of cell-substrate interactions – elastic force coupling in FAM vs. viscous drag coupling in HRM. Hence, by studying the mechanical interactions of motile cells it may be possible to distinguish between the two mechanisms of gliding motility. We tested the effect of mechanical forces on motility in two ways: (i) by examining the outcome of physical collisions between moving cells, and (ii) by probing the effect of applied load to the motion of individual motor complexes.

We hypothesize that the outcomes of mechanical collisions between a pair of cells will be different in the two models of motility because of the differences of the nature of cell-substrate interaction (see Fig. 1C). Specifically, during a cell-cell collision FAM-based cell motility would offer high resistance to the cell displacement because of the adhesive attachment between the cell and substrate. In contrast, HRM-based cell motility would result in a larger cell displacement, as the resistance due solely to viscous interactions would be weak.

To test this hypothesis, we built a computational model (see Methods and SI text for details) that represents the biophysical characteristics of *M. xanthus* cells and used it to investigate how the outcomes of cell-cell collisions depend on the gliding motility mechanism. Since the individual components and their interactions are not completely known at present, we simplified the two mechanisms of gliding motility in our model to focus exclusively on their cell-substrate interactions. As such we employed the viscous coupling model (VCM), which is similar the HRM and the elastic coupling model (ECM), which is similar to the FAM. We analyzed the modeling results for both mechanisms of cell - substrate interactions and compared them with quantified experimental data on isolated cell-cell collisions. Furthermore, we investigated the robustness of our results to variations in mechanical and geometrical parameters of the collision events.

As an additional test of the coupling of motors to the substrate, we used optically trapped beads to directly test the mechanics of motor coupling and the effect of load on motor movement. While the details of how applied hindering load affects the speed of the gliding motors themselves remains unknown, the two models of motility make qualitatively different predictions near the motor stall force due to the difference in coupling. Regardless of the shape of the motor force-velocity relationship, as the force is increased to high enough levels, the VCM predicts that beads should cease motion at the stall force and move backwards for higher loads. In contrast, the ECM predicts that applied force should stop bead motion even for loads well above stall.

# Results

## Distinct cell-cell collision behaviors of two alternative gliding motility models

To study the mechanical cellular interactions, we simulated a head-to-side collision between two cells moving on crossing paths. To differentiate the two mechanisms of gliding motility, we assumed strong attachments between the cell and the substrate in the ECM. The results show qualitatively distinct interaction behaviors of cells for the two alternative mechanisms (Fig. 2A and 2B). In these simulations, we define a primary cell as the one whose side is hit by the first node of another (secondary) cell. We observed that in the simulations of both the mechanisms, primary and secondary cells align with each other and move in a common direction after the collision. However, this common direction differed in the two mechanisms. In the VCM (Fig. 2A), both the cells changed their direction after the collision. In contrast, in the ECM (Fig. 2B), the new common direction is the same as the direction of the primary cell before collision. Thus, the primary cell maintained its direction, whereas the secondary cell aligned with the former.

This contrasting cell-cell collision behavior in the two mechanisms can be explained by observing the cell's resistance to shape deformation (bending). During collision the primary cell nodes are displaced due to the contact with the secondary cell, thereby leading to local deformation of the primary cell. This deformation results in counterforces on the cell nodes. In both of the mechanisms cell deformation produces viscous drag forces and angular spring forces on nodes. In addition, in case of the ECM a strong restoring force acts on nodes due to the substrate attachments. These forces do not allow the deformations to propagate to the front nodes of the primary cell and as a result limits its change in direction. Since no such force exists in VCM, cell-cell collision results in cell deformation that propagates to its front node, and in turn significantly changes the cell travel direction.

To identify which of the two scenarios resembles the behavior of colliding *M. xanthus* cells, we examined similar collisions in the time-lapse images of wild-type *M. xanthus* cells under low cell density ($8 \times 10^7$ cells/ml). We chose these conditions to easily identify isolated cells and their pairwise interactions. Figure 2C shows a typical cell-cell collision observed in experiments. In this case, the direction of the primary cell has not changed after collision. This behavior is similar to the simulations using the ECM of gliding motility. Thus, comparison of our simulations with experimental observations indicates there is an elastic coupling between the cell and substrate for *M. xanthus* gliding cells. Nevertheless, these conclusions may be sensitive to the parameter values used in our simulations or to the particular collision geometry. We therefore examined the robustness of these results.

## Distinct cell-cell collision behaviors require strong adhesion strength of substrate attachments but are robust to variation in other parameters

To further investigate the role of substrate adhesions during cell-cell collision we needed a quantitative metric to characterize the outcome of collision events. Since the major difference between the two mechanisms is the change in the primary cell orientation during collision, we focused on this value (see SI for details). We note that the collision outcome greatly depends on some aspects of the collision geometry, especially the collision position (defined from leading end of the cell, see Fig. S3A) and collision angle. Therefore, we set the collision angle as ~90 deg (that produces maximum change in cell orientation) and choose the maximum change in cell orientation ($\Delta\theta_{p,\max}$) from all possible node collision positions ($n = 2, 3, ..., N-1; N = 7$) as the metric that describes the cell-cell collision behavior for a specific parameter set. Nodes 1 and $N$ were excluded from this analysis since we assume no adhesion complexes at these nodes (see SI for details). Figure 3A depicts how $\Delta\theta_{p,\max}$ varies in the ECM model as a function of attachment strength. Each adhesive attachment is modeled by an elastic spring with a spring constant $k_a$ and a bond-breaking distance $L_{\max}$. By keeping the bond-breaking distance constant we vary the elastic spring constant ($k_a$) and thereby change the maximal force to break the bond. $k_a = 0$ corresponds to the case in which no bond is formed with the substrate, and therefore only viscous interactions with substrate exist (VCM). Intermediate values of $k_a$ correspond to weak and non-specific interactions with substrate therefore may correspond to viscoelastic properties of EPS slime surrounding cells. Large values of $k_a$ correspond to a strong specific binding which resembles the ECM. Figure 3A shows that the value of $\Delta\theta_{p,\max}$ starts at approximately 40 degrees for $k_a = 0$ and then decreases to values below 15 degrees for $k_a > 100$ pN/µm. $k_a = 100$ pN/µm corresponds to a bond-breaking force of 50 pN ($L_{\max} = 0.5\mu m$), which is roughly the order of integrin bond-breaking forces (~50-250 pN) (15, 16). This behavior is expected as the adhesion complex force will only be relevant if it resists the force generated by cell motility (~60 pN). When the attachment strength exceeds this threshold it results in small node displacements and hence small changes in cell orientations after collision.

For comparison we quantified the change in primary cell orientation ($\theta_{data}$) in isolated cell collisions under experimental conditions for 97 cell pairs (Fig. 3D). As these measurements also contain the spontaneous change in the orientation of cells, we measured the mean spontaneous orientation change ($\bar{\theta}_{basal} \sim 12\,\text{deg}$) of isolated cells (measurements from ~ 50 cells; see SI and Figure 3F and S3 for details) and subtracted it from experimental data ($\theta_{data}$). The mean value and standard deviation of the net change in cell orientation ($|\theta_{data} - \bar{\theta}_{basal}|$) from experimental images is shown in Figure 3A (gray area). Since these experimental results were based on wild-type cells that exhibit both gliding and twitching motility, we replicated the analysis with DK10407 ($\Delta pilA$ $A^+S^-$) cells that exhibit only gliding motility. The results from 58 isolated collision events for DK10407 cells are shown in Figure 3E. We observed that the mean change in primary cell orientation in isolated cell collision events for pure gliding motility cells is ~14±12 deg (n=58). This value is very close to that of the wild-type cells (~15±15 deg). These results indicate that the contribution of twitching motility to our analysis of gliding motility is negligible which is expected as twitching pili are located only at the poles and should not affect cell bending. Thus we used only wild-type cell data in our further analysis.

We observe that the modeling results match with the experimental observations only for adhesion strength ($k_a$) values greater than 200 pN/μm (Fig. 3A). Whereas, model results with no adhesion complexes ($k_a = 0$, VCM) show very large changes in cell orientation and do not match the experimental observations for the chosen parameters. However, we are uncertain whether the results will hold if some mechanical parameters (see Table S1 for model parameters) of the model are changed. We therefore systematically varied the model parameters over two orders of magnitude (from 0.1x to 10x, see Table S2) and investigated their effect on cell-cell collisions. These results are combined in Figure 4B, which shows the mean values and standard deviations of the cell orientation changes for different adhesion strengths ($k_a$). As before, we find that only the ECM model at very high adhesion strength values ($k_a > 500\,\text{pN/μm}$) matches the experimental values. We also noted that despite the large variability of the parameters, the standard deviations in the model results at high adhesion strengths are quite small. Thus, we conclude that our results are robust to variation in all model parameters except the strength of adhesion complex.

We have also quantified the minimum strength ($F_{min}^{adh}$) required for a focal adhesion complex in the cell model to match the experimental cell orientation change in a collision for different cell propulsive force ($F^p$) (Fig. 3C). The results indicate that $F_{min}^{adh}$ values increased with an increase in the cell propulsive force and are similar in magnitude to that of propulsive force.

## Distinct cell-cell collision behaviors are observed using the two gliding motility mechanisms over a range of collision geometries

As noted earlier, the outcome of cell-cell collisions depend on the mechanical parameters of the cell and on the collision geometry, specifically the collision position along the cell length and collision angle (the angle between the cell orientations at the start of the collision) (Fig. 4, S4).

To this point, our analysis focused on the maximum change in cell orientation as we varied the position of the colliding nodes and used a collision angle near 90 degrees at which the maximal cell deformation is expected. However, it is not clear if the experimental collision events correspond to these amplified effect scenarios or whether it is possible that a model with very weak or no elastic coupling can be consistent with the experimental collisions at some conditions. Thus, we systematically explored how variability in the geometrical model parameters affects the outcome of the collision. In these simulations we chose for the adhesion strength value ($k_a$) of 2000 pN/µm in the ECM, a value for which the model results closely match the experimental observations. In addition, all the experimental observations were corrected for spontaneous orientation change of cells ($\bar{\theta}_{basal}$).

First, we compared the orientation change from the two mechanisms as a function of the collision position for the default parameter set (Fig. 4A). We note that both models produced a much smaller change in cell orientations for collisions near the lagging end of the primary cell. This is an expected behavior of the model, as small node displacement near the lagging end of the cell may not produce sufficient cell deformation to significantly change the cell travel direction. However, collisions in the forward and middle section of the cell produced significantly larger orientation changes for the VCM model as compared to the ECM model. We note that the collision at the first node of primary cell produced very large orientation change in both mechanisms. This large change is due to the assumption that no adhesive attachment present at the first node of the cell (see SI for details). As a consequence we observed a large displacement of nodes even in the ECM. Thus, we excluded the first node collisions from our analysis. For comparison, we next quantified the changes in cell orientation as a function of the collision position from experiments (black circles in Fig. 3A). We note that only the results of the ECM model match with the mean experimental values for all collision node positions, whereas the results of the VCM model deviated significantly from the experimental values. We also found that these results are also robust to variation in mechanical parameters of the model (see Fig. S5A) and for small perturbations in collision positions (see Fig. S5B).

Next we investigated the effect of collision angle on cell-cell collision behavior. We varied the collision angle between 15 – 165 deg (corresponding to the experimental data, see Fig. S4C) and measured the maximum orientation change of the cell across all node collision position at each collision angle. We observed that the cell orientation changes with both the mechanisms are similar at both extremes of the collision angle range, but vary significantly in the middle (Fig. 4B). We compared these results with the observations from experimental cell collisions as a function of collision angle. We determined that results from the ECM model match closely with the experimental observations, whereas the results from the VCM model deviated significantly (Fig. 4B).

We also observed similar results for variation in cell length and number of adhesion complexes per cell (see Fig. S4C,D).Thus, the results from the VCM model consistently showed large cell orientation changes compared with the ECM model for various collision geometries. Further, the results from the ECM model match with mean values from the experimental data for all the collision scenarios considered.

## The effect of force on motor velocity is consistent with an elastic motor-substrate coupling

To directly test the coupling of single motor complexes to external objects such as the gliding substrate, we applied controlled loads to micron-sized beads being transported by gliding motors on immobilized cells (11). We designed a transient force clamp that isolates the effects of force on the motor-driven 'runs' even from the complex pause dynamics and occasional directional reversals seen in bead motion (Fig. 5A, see (11) and SI for experimental procedure). This procedure uses an optical trap to apply fixed loads to beads, but only after being triggered by a motor-driven displacement of 63 nm in less than 3 s. Trap position feedback was then used to maintain a constant force on the bead for approximately 8 s after which the trap was shut off. If the bead velocity and direction before and after force application was nearly the same, we concluded that the motor did not reverse or pause during the force-clamped period.

Fig. 6B-D show the measurements of bead linear velocities under various loading conditions in force clamp experiments. We observe that after some initial period of inactivity bead starts moving (green lines) at which point a preset opposing force is applied on the bead. We found that opposing forces of ~12pN (Fig. 5C) causes stalling of the bead whereas for forces below 12pN (Fig. 5B) bead movement is continued albeit slower than load free conditions. We estimated this stall force by finding those events in which the linear velocity was zero within twice the standard error of the linear velocity measurement. Interestingly, beads remained motionless for loads well-beyond the stall force (18 pN, Fig. 5D) and we never observed a bead to reverse its direction in response to high loads over the eight seconds of force application. The lack of backwards motion at super-stall forces is consistent with the ECM, but inconsistent with the viscously-coupled VCM model which predicts significant backwards motion at these loads.

We measured the force-velocity response of at least 108 motor complexes on 40 different *M. xanthus* cells. We chose the preset force to probe the complete force-velocity relationship for opposing loads from 0 to 20 pN (Fig. 5E) and also varied the concentration of nigericin (a drug that reduces pH gradient/ proton motive force across cell membrane there by decreases the motor function/bead motion (11)) in solution. We find that opposing loads slow gliding motors exponentially with a characteristic decay force of $2.3 \pm 0.1$ pN. In addition, we find that with increasing nigericin concentration, bead velocity decreased but force production did not (Fig. S6A-C). When normalized by the unloaded velocity, force-velocity curves from different nigericin concentrations collapse onto a single exponential curve (Fig. S6D) with characteristic force independent of nigericin. This is again inconsistent with the VCM in which decrease in velocity would lead to decrease in force production.

## Discussion

Despite progress in elucidating the mechanism of *M. xanthus* gliding motility, its biophysical mechanism is still not fully understood. Based on recent experimental evidence two alternative mechanisms: FAM and HRM of gliding motility are proposed but to date neither model has been conclusively proven. A key difference between the two models is in the biophysics of the interactions between cells and substrate. We hypothesized that this difference will affect cell behavior during cell-cell collisions. To test this, we constructed mathematical models of the *M. xanthus* cell with either viscous (VCM) or elastic (ECM) interactions with substrate and studied the mechanical behavior in isolated cell-cell collision events. As expected, we found that both models differed in their cell interaction outcome, which was quantified by cell orientation changes. We compared the results from both the models with experimental observations of isolated cell-cell collisions events under similar conditions. We found that experimental cell behavior differs from that of the VCM model and agrees with the ECM model in which there is strong adhesion between the cell and substrate. Variations of the mechanical and geometrical parameters in the cell model for the collision process further confirmed these findings and indicated the robustness of the model. Thus our analysis predicts strong elastic attachments between the cell and substrate, which is consistent with a focal adhesion mechanism for gliding motility. As a further test of the mechanics of cell attachment, we then studied of the effect of load on motor attachment and speed. We found that motors stalled to zero speed for loads about12 pN. Even when the load exceeded these stall force value (up to 20pN), the beads remained strongly attached to the cells and did not show motion in the opposite direction. This behavior is expected in an elastic-coupling model. In total, our simulations and measurements are consistent with the ECM and inconsistent with the VCM.

The strong attachment between cell and substrate indicated by our analysis are realistic and are similar in the range of other biological cell-substrate interactions (e.g. integrin focal adhesions in eukaryotic cells (15, 16)). Further, we observed that the minimum adhesive strength per node required to match the experimental observations increased with an increase in cell propulsive force, but remained within same magnitude (0.5x – 5x) of cell propulsive force. Based on the force-clamp experimental estimate of ~12pN force generated at each focal adhesion node, and assuming ~5 adhesion nodes per cell (10, 17), we estimate the gliding motility apparatus generates ~60pN of force. This estimate is of the same order as the force generated by the twitching motility engine (18, 19) which is not surprising given that *M. xanthus* cells using either gliding or twitching motility move at approximately the same speed (20). In light of this estimate, our model (Fig. 3C) would predict at least 80pN of adhesion force. In support of this, force-clamp experiments never observed bead detachments for forces up to 20pN.

While the work here probed the attachment of motor bound-bead cargos to immobilized cells, it should be possible in the future to directly test the ECM model using optical tweezers and moving cells to measure the cell detachment force along the cell length. Large detachment forces with the existence of multiple peaks in the cell displacement curve along the cell length would provide support for multiple strong attachment sites.

Although our biophysical model that includes strong adhesion, similar to the FAM, explains the observed experimental cell collision behavior, a number of issues remain unresolved regarding the focal adhesion model of gliding motility. First, while it is observed that clusters of AlgZ proteins, which are predicted to form the focal adhesions, remain stationary during cell movement (5, 10, 21), this behavior requires that the adhesion complexes move through the peptidoglycan layer. Second, the adhesive proteins/molecules that bind the motor complexes to the substrate have not been identified. A recent study by Durcet et al. (22) speculated that slime acts as a binding agent between the cell and substrate. In this context it is worth noting that our biophysical cell model incorporates a simplistic viscoelastic model for cell-substrate interactions. However, a non-isotropic viscoelastic model for attachment may provide a better description of the substrate interactions (23). Third, strong cell-substrate attachments pose an additional problem for cell by restricting its movement at the lagging pole. Since the attachments remain stationary during cell movement, the elastic nature of the attachment at the lagging pole would be expected to cause an increasingly opposing force for cell movement as the cell moves forward until the attachment is broken causing its lagging pole to snap back. This type of jerky motion is commonly seen in fibroblasts that utilize substrate attaching lamellopods for movement (24-26). However, since this type of motion is absent in *M. xanthus* gliding, it suggests that the cells actively destroy attachment complexes at the lagging pole.

Is there any physiological role for the strong adhesion with substrate? We speculate that the strong attachment between the cell and substrate helps the cells align at high cell density. Indeed, the simulations of Janulevicius et al. (27) lacking substrate adhesion, indicated that *M. xanthus* cells with the bending modulus reported in literature (28, 29) cannot maintain alignment. In our model we have observed that when strong substrate adhesion is included the orientation of one of the cells remains unchanged during cell collisions, whereas the orientation of the other aligns to this orientation. This reflects the natural arrangements of high density *M. xanthus* cells that self-organize into well-aligned clusters (30, 31). As new cells join and align with the existing cells in a cluster, strong substrate attachments prevents the change in orientation of the cell clusters, thus preserving the mean orientation of the cluster. This effect appears to explain that flexible cells can maintain their alignment using strong adhesive attachment with the substrate.

# Methods

To simulate the cell motility and cell-cell collisions, we have developed a biophysical model of *M. xanthus* cell by extending the linear flexible cell model(27). In our model, each cell is represented as a connected string of nodes. Neighbor nodes are joined using linear and angular springs that simulate elastic behavior of the cell. Cell propulsive forces are applied at the center of nodes to simulate force generation from motor protein complexes. Cell experiences drag force on its nodes that oppose cell movement as it travels on the slime. Cell nodes interact with the substrate using elastic attachments (in ECM) modeled as linear springs that resist displacement of nodes during cell-cell collision. These attachments are absent in VCM. Collision resolving forces are applied on the nodes when nodes are in direct contact. We solve the equation of motion describing the cell movement by integrating the Newton's laws of motion for all the nodes using an open source physics library Box2D (http://box2d.org/).

To compare our model results with collision behavior of actual cells, we have analyzed similar cell-cell collisions in time-lapse images of *M. xanthus* cells (DK1622, DK10407) obtained under low cell density ($8 \times 10^7$ cells/ml) conditions. For imaging, cells are grown in CTT broth and then are diluted, spotted on agar plate and allowed to acclimate for at least 2 hrs before imaging with an Olympus 81X inverted microscope fitted with a Hamamatsu HD camera. The cells were imaged at 5 or 10 sec intervals for up to 12 hrs. We have tracked individual cell positions and orientations in colliding cell pairs from time-lapse images using ImageJ (32) software and measured change in primary cell orientation for each collision event. We corrected these measurements by measuring orientation changes of isolated cells.

To quantify the mechanics of the interaction between gliding motors and the substrate we have measured the velocities of optically trapped beads moving on the cell surface in force-clamp experiments (11). Cells (DZ2) are immobilized in custom-made flow chambers (33). Polystyrene beads (diameter 520 nm) are then optically trapped and placed on the top of the cells. Bead position is detected using a low-power diode laser (wavelength 855 nm) aligned with the optical trap. The microscope stage is adjusted in 3D to insure the bead remains in the center of the detector laser using a PID feedback algorithm. During force clamp operation, the optical trap is moved to an off-center position with respect to the detection/tracking laser along the cell axis, but with the shutter closed. If the bead moves at least 63 nm in one direction along the cell axis within a time interval of less than 3 s and without tracking backwards, the shutter opens, thus applying a preset force to the bead. The distance between the detection and trapping laser is kept constant, while feedback of the stage position keeps the bead centered in the detection laser. The trap is released after approximately 8 seconds. Runs with the same velocity before and after force application (within measurement error) are kept for analysis.

See SI for further details on modeling and experimental procedures.

# Acknowledgements

This work was supported by National Science Foundation CAREER awards to OAI (MCB-0845919) and JWS (PHY-0844466). The simulations were performed using Rice University cyber infrastructure supported by NSF Grants CNS-0821727 and OCI-0959097.

**Figure 1**

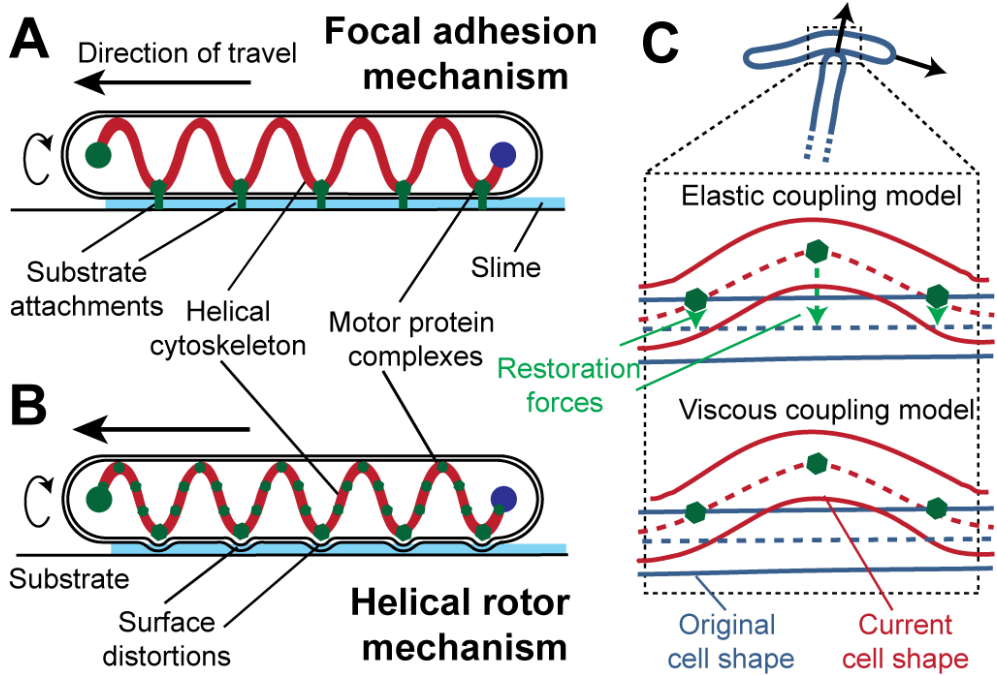

Figure 1: **Schematics of alternative mechanisms of gliding motility and their representation in biophysical models of the *M. xanthus* cell**. (A) Focal adhesion mechanism (FAM) - Multi-protein complexes (green bars) spanning from the cytoplasm to the outside of the cell attach to the underlying substrate at specific points. Cells move forward as a result of the force generated by the components of these complexes against cytoskeleton (B) Helical rotor mechanism (HRM) - Motor proteins (green dots) tracking on a helical cytoskeleton create distortions in cell wall. These distortions generate drag forces between the substrate and the cell surface and result in cell movement. (C) Distinctions in cell-substrate interactions for the two alternative models of gliding motility. In the elastic coupling model during a cell-cell collision, a restoration force acts on the cell at the cell-substrate interaction points (green dots) in the direction perpendicular to cell axis. No such force exists in the viscous coupling model.

**Figure 2**

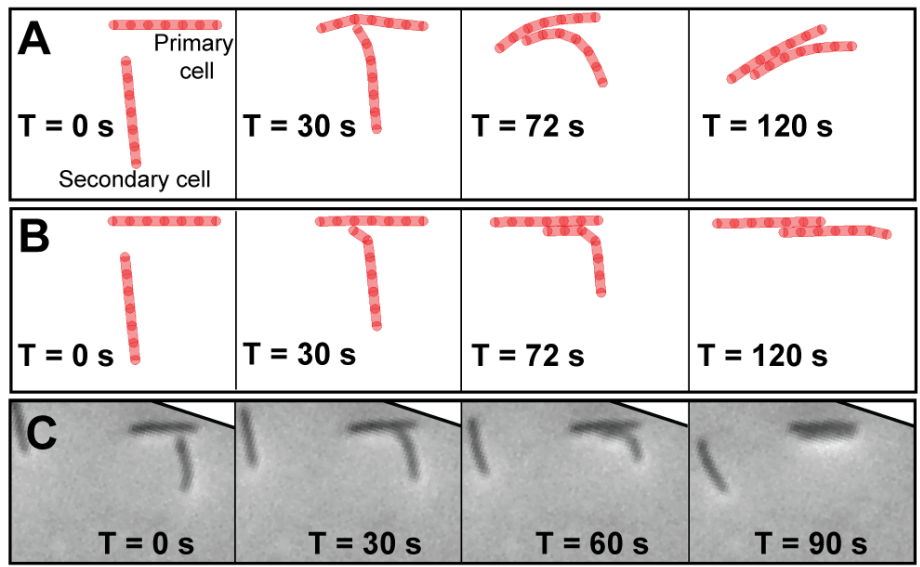

Figure 2: **Mechanical interactions between two cells during head-to-side collisions in the biophysical models and experiments**. (A) Viscous coupling model – both cells change directions. (B) Elastic coupling model – only the secondary cell changes direction. (C) Experimental time-lapse images (rotated to match with simulation configuration) showing collision between two isolated cells where only the secondary cell changes its direction.

**Figure 3**

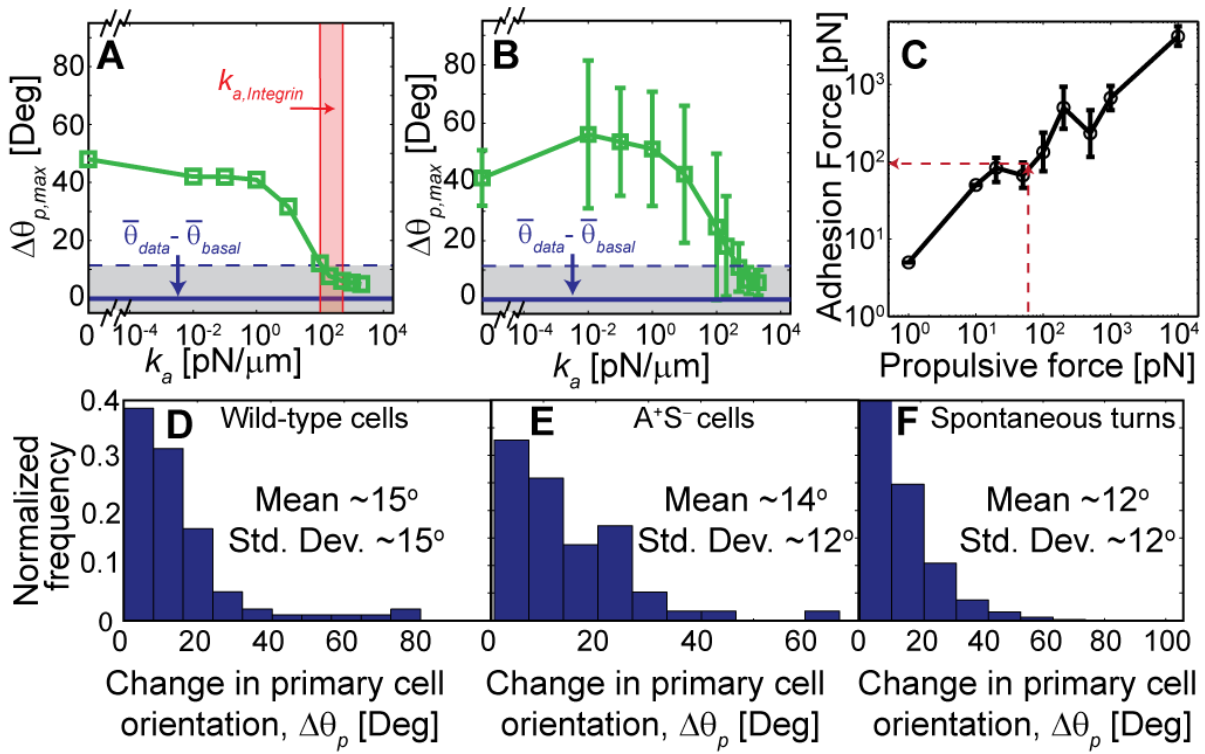

Figure 3: **Strong adhesive attachments between cell and substrate are required to match experimental observations.** (A) Maximum change in primary cell orientation ($\Delta\theta_{p,\max}$) as a function of the strength of substrate attachments ($k_a$). Red band represents the range of bond strengths observed for integrin bonds in other biological systems (22-24). Horizontal solid line ($\left|\overline{\theta}_{data} - \overline{\theta}_{basal}\right|$) represents the mean value of change in primary cell orientation from experimental cell collisions after subtracting the spontaneous cell turning and the dashed lines represent one standard deviation variation in the experimental data. (B) Same as (A) but with mean and standard deviation from aggregated simulations with varied model parameters. (C) The minimum adhesive strength of attachments matching experimental data closely matches with the cell propulsion force. Error bars represent variation in the results for different cell flexibilities. (D) The distribution of $\Delta\theta_p$ values in experimental data of wild-type cells (DK1622, collision events, N=97) and (E) cells lacking twitching motility (DK10407, N=58). (F) The distribution of spontaneous cell orientation change for mean cell collision time of ~2.9 min measured from trajectories of isolated cells (DK1622, N=4018, see Fig. S4A for additional details).

**Figure 4**

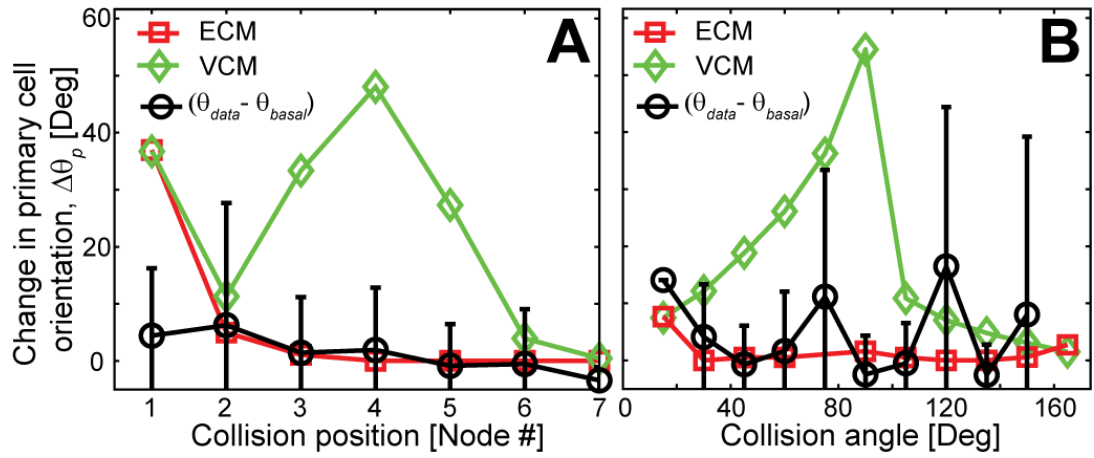

Figure 4: **Distinct cell behavior from the two cell models for variation in collision geometries.** (A) The change in primary cell orientation ($\Delta \theta_p$) as a function of collision position from the leading end of the primary cell and (B) as a function of collision angle.

**Figure 5**

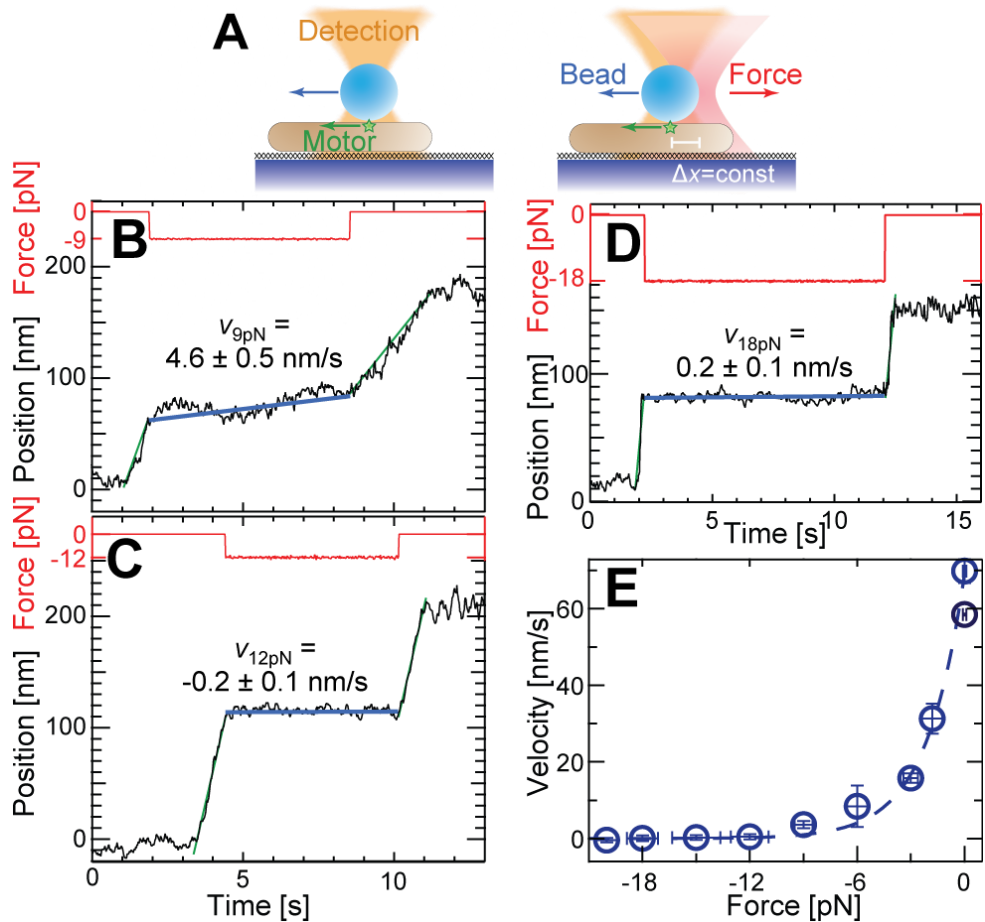

Figure 5: **Bead/molecular motor motility behavior under optical trap loading.** (A) A gliding motor moves a bead along the cell axis. Past a preset threshold movement, the shutter in front of the optical trap is opened, pulling the bead in the direction opposite to the motor by the preset force, resulting in a slowing of bead movement. (B-D) For opposing forces of 12 pN or greater bead movement has stopped and for lower forces (9 pN) bead movement is slowed down but not completely stopped. Here an experiment is associated with the activity of a single motor only if the bead moves before and after trapping with the same direction and speed (green lines). A linear fit to the position versus time during force application provides the velocity (blue lines). (E) Bead velocity decreases exponentially with force but never becomes negative consistent with an elastic coupling and inconsistent with a viscous coupling between the bead and motor. The dashed lines are an exponential fit to the data. Error bars represent the standard error of the mean across trials (> 6 trials per data point).

# Supplemental Text and figures

## Computational methods - Biophysical model of cell motility

We developed a biophysical model of the *M. xanthus* cell by extending the linear flexible cell model by Janulevicius et al.(1). In our model, each cell is represented as a connected string of $N$ circular nodes with a total cell length $L$ and width $W$ (Fig. S1A). Each circular node is modeled as a rigid body of radius $W/2$. Circular nodes are kept at a fixed distance apart by $M(=N-1)$ rectangular spacers of length $((L-W)/(N-1))$ and width ($W$). Each body of mass ($m_i$) is identified by its position ($r_i$) and heading direction ($\theta_i$). Neighbor circular and rectangular bodies are connected by a rotational joint at the center of the circular node (Fig. S1B). Each circular node is connected to the neighboring circular nodes by angular springs ($a_i$) that resist bending of the nodes from straight line position (Fig. S1C). Spring constants for the angular springs ($k_b$) can be tuned to achieve the desired flexibility of the cell that matches with the actual bacterium. Various forces ($F_i$) (e.g. propulsive forces that move cell forward, drag forces on the cell surface due to contact with surrounding fluid) act on the nodes that affect the velocity ($v_i$) of the nodes.

In the following sections we describe the equations that model the cell motion. In these equations letters represent magnitudes and bold letters represent vectors.

### *Rotational joints*

Rotational joints between circular and rectangular bodies are modeled as linear springs with zero equilibrium length (Fig. S1B). Thus, joints resist variation in length (elongation and compression) between connected bodies with counteracting forces ($F_i^l$) determined by Hooke's law

$$F_i^{l,i} = -k_l l_i = -F_{i+1}^{l,i}$$

where $l_i$ is the vector joining the connected bodies at joint $i$ from their respective joint positions ($|l_i| = 0$ at equilibrium) and $k_l$ is linear spring constant.

### *Angular springs*

Angular springs resist bending of the bacterium to simulate elastic behavior of the cell. An angular spring $a_i$ connects every three adjacent circular nodes $i, i+1$ and $i+2$ (Fig. S1C), where $i = 1, 2, \ldots$, N-2. Each angular spring exerts torques ($\tau_i^{a,i}$ and $\tau_{i+2}^{a,i}$) on the connected arms ($p_i$ and $p_{i+1}$) of the spring.

$$\tau^{a,i} = k_b \theta_i$$
$$\tau_i^{a,i} = \left( (p_{i+1} \times p_i) / |p_{i+1} \times p_i| \right) \tau^{a,i}$$

where, $k_b$ is the angular spring constant, $\theta_i$ is the angle between the nodes, $\tau_i^{a,i}$ is the torque acting on the node $i$, $p_i$ is the vector joining the neighboring circular nodes $i+1$ to $i$. These torques are converted to forces ($F_i^a$) on nodes such that forces acting on node $i$ and $i+2$ cancel out the force acting on node $i+1$ thus producing zero axial movement of the nodes.

$$F_i^{a,i} = (p_i \times \tau_i^{a,i}) / p_i^2$$
$$F_{i+2}^{a,i} = (p_{i+1} \times \tau_{i+2}^{a,i}) / p_{i+1}^2$$
$$F_{i+1}^{a,i} = -(F_i^{a,i} + F_{i+2}^{a,i})$$

*Cell motility forces*

Motility forces are the forces that are generated internally in the cell and propel it forward. In this model, we consider only the force generation due to gliding motility. In both the proposed mechanisms of gliding motility, cell propulsive force is generated at the motor protein complexes. In our model circular nodes are equivalent to these motor protein complexes. Thus, we apply motility/propulsive force ($F_i^p$) at each circular node $i$ (except first and last node) of the cell along the segment connecting next circular node in cell travel direction (Fig. S1D). We assume that force generated per node due to gliding motility is constant.

$F_i^p = \dfrac{p_i}{p_i} F_i^p$, where $p_i$ is the vector joining the neighboring circular nodes $i+1$ to $i$.

*Viscous drag forces*

*M. xanthus* cells secrete slime from their surface, which is deposited on the underlying substrate as long trails (2). As cells move on the slime they experience drag forces that oppose their movement. Since the mean speed of *M. xanthus* cells is very low (4 µm/min) (3) and the dimensions of the cell are ~ 0.5 × 7 µm (4), the cell movement is in the low Reynolds number flow regime, and thus we assume only a Stokes drag force acts on the cell.

Stokes drag force on body $i$ is determined using the equation $F_i^d = -c v_i$. Here, $c$ is the drag coefficient between body and slime and $v_i$ is the velocity of the body $i$. Drag coefficient is adjusted such that the terminal speed ($v_f$) achieved by the model cell based on the total force generated matches the mean speed of *M. xanthus* cells observed experimentally.

*Node-substrate interaction force*

Adhesive attachments between the cell and the substrate result in a restoration force ($F_i^r$) on circular node $i$ when the node is displaced from its position due to collision with another cell

(Fig. S1E). These forces restore the displaced node to its original position. Here the attachments are modeled as linear springs, with a spring constant $k_a$, that break if stretched beyond a threshold length ($L_{max}$).

$$F_i^r = \begin{cases} k_a r_i^d & \text{for } r_i^d \leq L_{max} \\ 0 & \text{for } r_i^d > L_{max} \end{cases}, i = 2, 3, ..., N-1$$

where $r_i^d$ is the perpendicular displacement of the node from its original position due to collision. Here we assume that the cell does not form an attachment at the first node as it interferes with the cell's ability to randomly change its direction, which is normally observed in *M. xanthus* cells. Node-substrate interaction forces are absent in VCM (Fig. 1C).

When an attachment is broken it reforms after a random waiting time ($\tau$) that is exponentially distributed with a mean of 1/8 min (rate of new bond formation of 8 1/min). The mean waiting time is estimated on the experimentally observed cell speed ($4\mu m/\min$) and pitch of helical cytoskeleton ($\sim 0.5\mu m$) (5). We assumed that the waiting time corresponds to the time for the arrival of the next motor protein to the next node.

*Collision forces*

When two cells collide (i.e. bodies/nodes of two cells are in direct contact) collision resolving forces ($F_i^c$) are applied on the nodes to stop them from overlapping (Fig. S1F). These forces are applied in the direction normal to their surfaces at the point of the collision. Collision detection and collision resolving forces are handled by a physics engine (see below) in our model. .

*Equations of motion*

The equations of motion that describe the movement of a cell in the model are as follows.

For each body $i$, total force $F_i^T = F_i^l + F_i^a + F_i^p + F_i^c + F_i^r + F_i^d$

Angular spring forces ($F_i^a$) and propulsive forces ($F_i^p$) are absent on rectangular bodies.

Positions of and velocities of nodes ($r_i$, $v_i$), are determined by integrating the Newton's laws of motion (shown below) using Box2D physics engine.

$$\frac{dv_i(t)}{dt} = \frac{F_i^T}{m_i} \text{ and } \frac{dr_i(t)}{dt} = v_i^f(t), i \in \{1, 2, ..., N+M\},$$

*Box2D physics engine*

We use an open source physics library Box2D (http://box2d.org/) to solve Newton's equations of motion in our model. Box2D is a two-dimensional rigid body dynamics simulator that solves the equations of motion of bodies subjected to various forces, and outputs the position and velocity of the bodies at each time step. We modeled the biophysical cell in Box2D, using the mathematical modeling approach described above. The model parameters were specified as various physical parameters to the simulation engine. The collision forces between the bodies were internally calculated by the physics engine. The cell model is simulated at each time step where the position, orientation and velocity of the nodes are recorded. The parameters used in the model are listed in Table S1. We scaled the actual cell parameters to the model cell configuration due to the restrictions on the rigid body dimensions that Box2D simulates. These restrictions are introduced primarily to maintain the numerical error within acceptable limits and for the numerical stability of the simulation. We have also modified the integration scheme used by Box2D to the semi-implicit Euler method from the original explicit Euler method.

*Quantifying cell-cell collision behavior*

To quantify the cell-cell collision behavior we used the following procedure. We numbered the nodes in each cell from the leading end (node $1$) to the lagging end (node $N$), and defined a vector ($O$) pointing from lagging to leading node as the travel direction of the cell. Orientation of cell '$i$' at any instance of time '$t$' is denoted by $\theta_i(t)$ and is quantified as the angle difference between the cell's travel direction vector ($O$) and the horizontal axis ($y=0$) in the counter-clockwise direction (Fig. S1A). We defined a primary cell as the one whose side is hit by the first node of another (secondary) cell. Thus, the change in orientation of both the primary and secondary cells can be recorded as a function of time. We measured the value of the primary cell orientation change ($\Delta\theta_p = |\theta_{t(+\infty)} - \theta_{t(-\infty)}|$) before and after the collision. A schematic showing the quantification of cell collision process and the corresponding change in cell orientations are shown in Figures S2B and S2C.

We observed that the change in the primary cell orientation during the collision process varies based on the collision position (node numbers) along its length (Fig. S3A and Fig. S3B). We used the maximum change in cell orientation ($\Delta\theta_{p,\max}$) resulting from all possible node collision positions ($n=2,3,...,N-1$) as a metric to compare the model results.

*Quantifying cell collision behavior from experimental time-lapse images*

Cell-cell collision behaviors under experimental conditions were quantified by tracking the cell's position and orientation in time-lapse images during the collision process. We used ImageJ software (6) with MTrackJ plugin (7) for cell tracking. First, we identified cell collision events that were free from interactions with neighboring cells (isolated collision events) in the experimental time-lapse images. Next, we loaded the image stacks corresponding to the collision events from the time-lapse movies into ImageJ. We used MTrackJ (a cell tracking plugin for ImageJ) to track the positions of the colliding cells as a function of time. The individual cell's leading and lagging ends were marked manually for each image in the loaded frame stack. These marked positions were converted to a time-series of $(x, y)$ pixel coordinates by MTrackJ (Fig. S2C). From the pixel coordinates (the cell's leading and lagging pole positions) we calculated the cell orientations as a function of time (Fig. S2D, see previous section for details).

*Measuring the spontaneous turning of M. xanthus cells*

The change in the primary cell orientation that we measured in our experimental time-lapse images also includes an additional component due to the spontaneous turning of the cells. To estimate the actual change in cell orientation due to collision, we measured the mean orientation change ($\bar{\theta}_{basal}$) of isolated cells for the duration of mean collision time ($\bar{t}_c$) (Fig. 3E, S4A). We subtracted this value from our experimental estimates of cell orientation change. To measure the mean orientation change of individual cells, we first tracked the orientation ($\theta_i(t)$) of isolated cells over time (Fig. S4B) from the time-lapse images and then calculated the mean cell orientation change using the following equation.

$$\bar{\theta}_{basal}(\bar{t}_c) = \frac{1}{N \times K} \sum_{i=1}^{N} \sum_{t=0}^{T} \left| \theta_i(t) - \theta_i(t + \bar{t}_c) \right|$$

where $N$ is the total number of cells tracked, $\left| \theta_i(t) - \theta_i(t + \bar{t}_c) \right|$ is the absolute orientation change of cell $i$ in the time interval $\bar{t}_c$, and $K$ is the number of such possible measurements for a total tracking time of $T$.

Table S1: Parameters used in flexible cell model

| Symbol | Description | Value |
|---|---|---|
| $L$ | Cell length | 6.5 µm (1, 8) |
| $W$ | Cell width | 0.5 µm (1, 8) |
| $N$ | Number of nodes per cell | 7 |
| $N_a$ | Number of adhesion complexes | 5 (9, 10) |
| $\rho$ | Cell density | 1000 kg/m$^3$ |
| $k_l$ | Linear spring constant | Managed by Box2D |
| $k_b$ | Angular spring constant/bending stiffness | 10$^{-15}$ N.m (1, 11, 12) |
| $k_a$ | Substrate attachment spring constant | 50-2000 pN/µm (13, 14) |
| $L_{max}$ | Bond breaking length | 0.5 µm (cell width) |
| $v^f$ | Mean speed of the cell | 4 µm/min (4) |
| $t_{step}$ | Simulation time step | 5×10$^{-3}$ sec |
| $\mu$ | Viscosity of the slime | 10$^{-3}$ kg/m.s |
| $F^p$ | Propulsive force per cell | 60 pN |

Table S2: Mechanical parameters varied in the model for testing the robustness of model results

| Parameter | Range varied |
|---|---|
| Angular spring constant ($k_b$) – represents cell flexibility (1, 12) | 10$^{-14}$ – 10$^{-16}$ N.m |
| Spring constant of substrate attachment ($k_a$) | 0 – 2000 pN/µm |
| Drag coefficient between cell surface and substrate environment | 9×10$^{12}$ – 9×10$^{14}$ kg/s |
| Bond breaking length ($L_{max}$) | 0.25 – 1.0 µm |

# Experimental methods

## Cell Growth and Development

For cell collision experiments *M. xanthus* strains DK1622 (wild-type) and DK10407 ($A^+S^-$) were grown in CTT broth (1% Difco Casitone, 10 mM Tris-HCl pH 8.0, 8 mM $MgSO_4$ and 1 mM $KHPO_4$ pH 7.6) or on CTT agar (CTT broth containing 1% agar) at 32°C. When *M. xanthus* cells reached mid-log phase ($4\times10^8$ cells/ml), the cells were diluted to 20% in TPM buffer (CTT without Casitone).

For optical trap experiments *M. xanthus* strain DZ2 AglZ-YFP ΔpilA (9) was incubated on 1.5% agar plates (CYE medium - 1% peptone, 0.5% yeast extract, 10 mM MOPS, pH 7.8) at 32°C for 4 days. 10 μl of cells were transferred in 25 ml CYE containing 10 mg/ml tetracycline. Cultures were incubated in a shaker at 32°C overnight. Prior to experiments, 2 ml cell culture grown to an OD of 0.78 was centrifuged at 8,000 rpm for 5 min, the supernatant removed, and pelet resuspended in 400 μl TPM medium (10 mM Tris, 8 mM $MgSO_4$, 100 mM $KH_2PO_4$, pH 7.8).

## Microscopic imaging of cell collisions

Corning 35 mm tissue culture dishes were prepared for microscopy by drilling a 5 mm hole in the bottom of the dish. A microscope cover slip was then taped over the hole and 5 mls of 1/2 CTT agar was poured into the culture dish. After the agar solidified, the microscope cover slip was removed and 5 μl of the diluted *M. xanthus* (DK1622, DK10407) culture was spotted onto the exposed agar and allowed to dry. The culture dish was then inverted and water was added to the large agar surface. The cells were allowed to acclimate for at least 2 hrs before imaging with an Olympus 81X inverted microscope fitted with a Hamamatsu HD camera. The cells were imaged at 5 or 10 sec intervals for up to 12 hrs. The temperature was maintained at 30°C during the imaging.

## Flow chamber and surface preparation for bead assays with immobilized cells

Flow chambers were custom-made using double layers of double-stick tape and a cover slide (thickness 1 mm) and a cover slip (#1.5, thickness 100 μm) as described previously (15). The final chamber volume was approximately 40 μl. 20 μl of 0.7% agarose dissolved in 6 M DMSO were injected into a chamber, incubated at room temperature for 15 min and washed with 400 μl TPM (16). *M. xanthus* (DZ2) in TPM was injected and allowed to attach firmly to the surface for 30 min. Non-attached cells were washed out thoroughly using 2 ml TPM containing 10 μM glucose. Samples were mounted onto the microscope and ready for experimental use.

For all bead experiments, polystyrene beads (diameter 520 nm) were washed and diluted in TPM medium (0.005% weight/volume) containing 10 μM glucose and injected into the flow chamber. Freely diffusing beads were optically trapped and placed on surface-immobilized cells. For subsequent experiments in the presence of different concentrations of the drug nigericin, TPM medium with the appropriate nigericin concentration was carefully injected into the microscope-mounted flow chamber.

*Optical tracking and trapping*

The optical trap was custom-built onto a Nikon TE2000 microscope equipped with a TIRF objective (NA = 1.49, Nikon). A trapping potential for transparent objects was formed by focusing the $TEM_{00}$ mode of a high-powered $Nd:YVO_4$ laser (wavelength 1,064 nm, up to 5W output power). A piezo-controlled tip-tilt mirror allowed for precise positioning of the optical trap within the focal plane. The flow chamber was mounted on top of a 3D-piezo stage with a wide working range (200 μm × 200 μm × 200 μm). The experiments were recorded by an EMCCD camera mounted behind a 2.5 times zoom with a field of view of 41 μm × 41 μm.

A low-powered diode laser (wavelength 855 nm, operated at 3 mW output power) was aligned with the optical trap, and the forward scattered light from trapped objects was collected onto a position-sensitive photodiode placed in a plane conjugate to the back aperture of the condenser, pre-amplified at the diode, amplified and filtered (low-pass filter 53 kHz), and recorded with a data acquisition card. Using this acquisition, we implemented a PID feedback to fix the bead position relative to the detector laser beam focus by moving the piezo. Typically, we recorded positional tracking of bead movement on the cell surface with 10 kHz using stage feedback in the lateral direction with a frequency of 50 Hz to keep the detection/tracking laser focused on the bead. The height was fedback at a frequency of 1 Hz by comparing the bead's image with a look-up table taken prior to the actual tracking. Tracking could be performed for hours without significant drift. We stored the images of the whole field of view at 1 Hz to ensure that no other objects diffused into the path of the detection laser close to the focus. In a post-processing step, the images were used to project the lateral dimensions of the bead trajectories along the cell major axis.

Prior to deposition onto the cell, trapped beads were calibrated by monitoring their Brownian motion at an acquisition frequency of 22 kHz for 10 s in close vicinity to the cell. A standard protocol was employed to extract the harmonic trap stiffness, the linear photodiode-voltage to position conversion factor, and an unbiased measure of the accuracy (17). The sum signal of the photodiode was used as a measure of bead displacement in the direction perpendicular to the focal plane.

For force-clamp experiments, the optical trap was moved to an off-center position with respect to the detection/tracking laser along the cell axis, but with the shutter closed. The exact distance was chosen under the assumption that the exerted force was linear to the displacement from the trap center (18). If the bead moved at least 63 nm in one direction along the cell axis within a time interval of less than 3 s and without tracking back, the shutter would open, thus applying the preset force of the optical trap to the bead. Updating the bead position effectively functioned as a force feedback, since the distance was kept constant at the same time. The trap was released after 8 seconds. Runs where the kept if the velocity before and after force application were the same within measurement error.

**Figure S1**

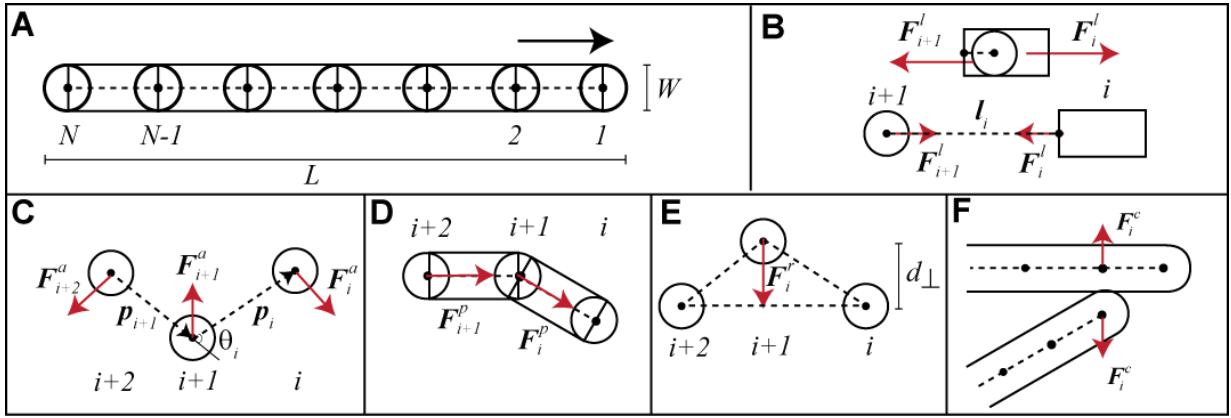

Figure S1: **A biophysical representation of the *M. xanthus* cell as a mass-spring system.** (A) Each flexible cell is represented as a connected string of nodes. Circular nodes are similar to the focal adhesion complexes (FAM) or cell wall distortions (HRM) at which cell propulsion force is generated. Circular nodes are spaced apart by rectangular bodies of fixed dimensions. (B) Linear springs join neighboring nodes (circular and rectangular bodies) and maintain the connectivity between the bodies by opposing change in distance between the connected bodies and apply forces ($F^l$) to that effect. (C) An angular spring between three consecutive circular nodes resists bending of the nodes from straight line formation by introducing elastic bending forces ($F^a$). (D) Cell propulsive forces are applied at circular nodes along the segment joining the next neighbor node in cell travel direction. (E) Adhesive attachments between the node and the substrate are represented by linear springs that introduce a restoration force ($F^r$) on that node when it is displaced from cell's linear axis. (F) Collision forces ($F^c$) act on the nodes that are in direct contact to prevent overlap of bodies.

**Figure S2**

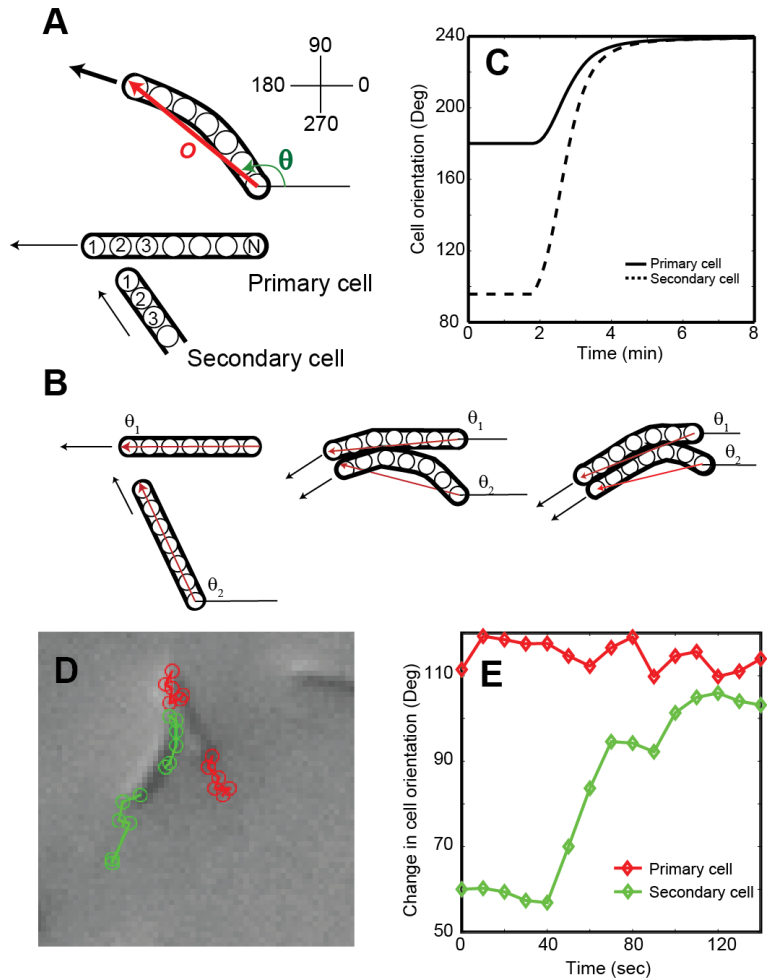

Figure S2: **Quantification of cell-cell collision behavioral data from simulations and experiments.** (A) A cell's travel direction is indicated by the vector ($O$ – red arrow) pointing from lagging to the leading end. This direction is converted into the cell orientation ($\theta$) by measuring the angle between the cell direction vector ($O$) and the horizontal axis (y=0) in the anti-clockwise direction. The position of cell collision is identified by the node number of the primary cell where the secondary cell first makes contact. (B) Schematics showing the change in cells' orientations at different instances of time during the collision process. (C) Simulation results show the corresponding change in cell orientations with time. (D) Cell tracking using ImageJ software and MTrackJ plugin. Individual cells (red, green) participating in collision are identified and their leading and lagging ends are marked in consecutive time-lapse images during collision process. The chain of points for each color represents the tracking history of a marked cell's end through the time-lapse images. (E) Orientation of primary (red) and secondary cell (green) as a function of time during a collision event as measured from time-lapse images. Observe that after the collision secondary cell (green line) changes its orientation and aligns with the primary cell (red line).

**Figure S3**

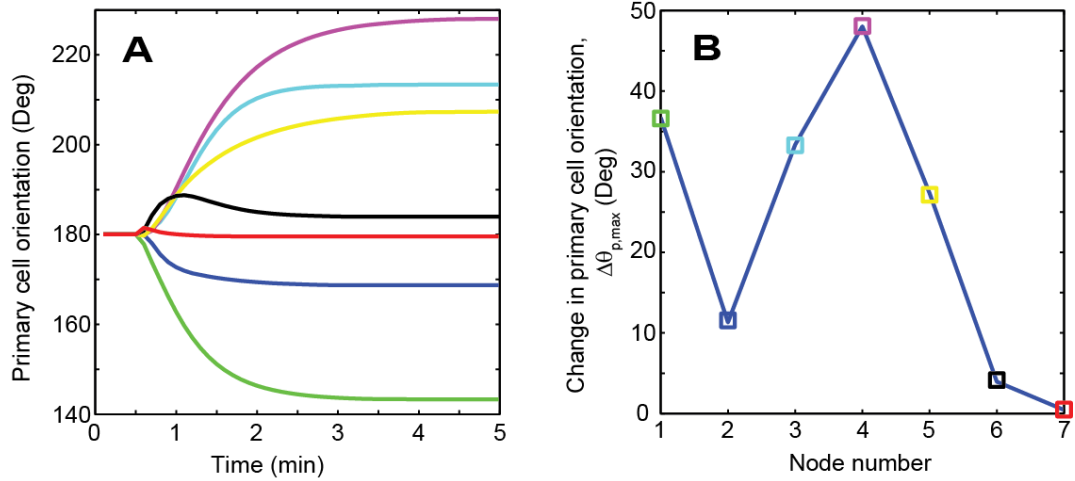

Figure S3: **Cell collision behavior varies for different collision positions** (A) Change in the primary cell orientation with time for different collision node positions (different colors) from the leading end of the primary cell. (Note that collisions occur around 2 min). (B) Absolute change in the primary cell orientation before and after the collision as a function of node position. Data points shown in different colors correspond to the lines in panel A.

**Figure S4**

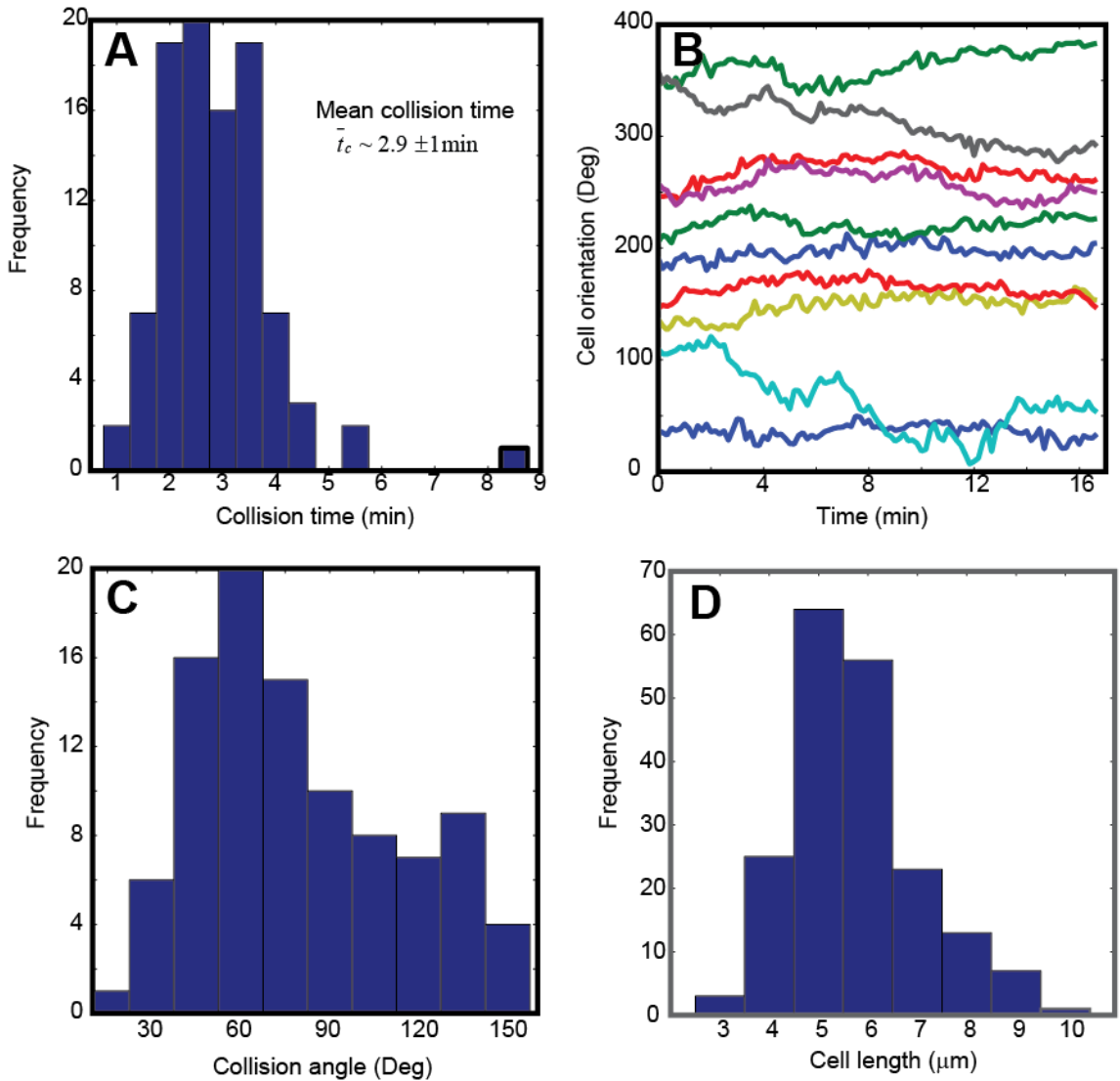

Figure S4: **Cell properties measured in experimental data for wild-type cells.** (A) Distribution of collision times from the experimental data of wild-type cells. (B) Tracking history of individual cell orientation over time indicating the spontaneous random cell orientation changes. Each trajectory/color represents measurements from a single isolated cell. (C) The distribution of collision angles and (D) cell lengths from wild-type cell data.

**Figure S5**

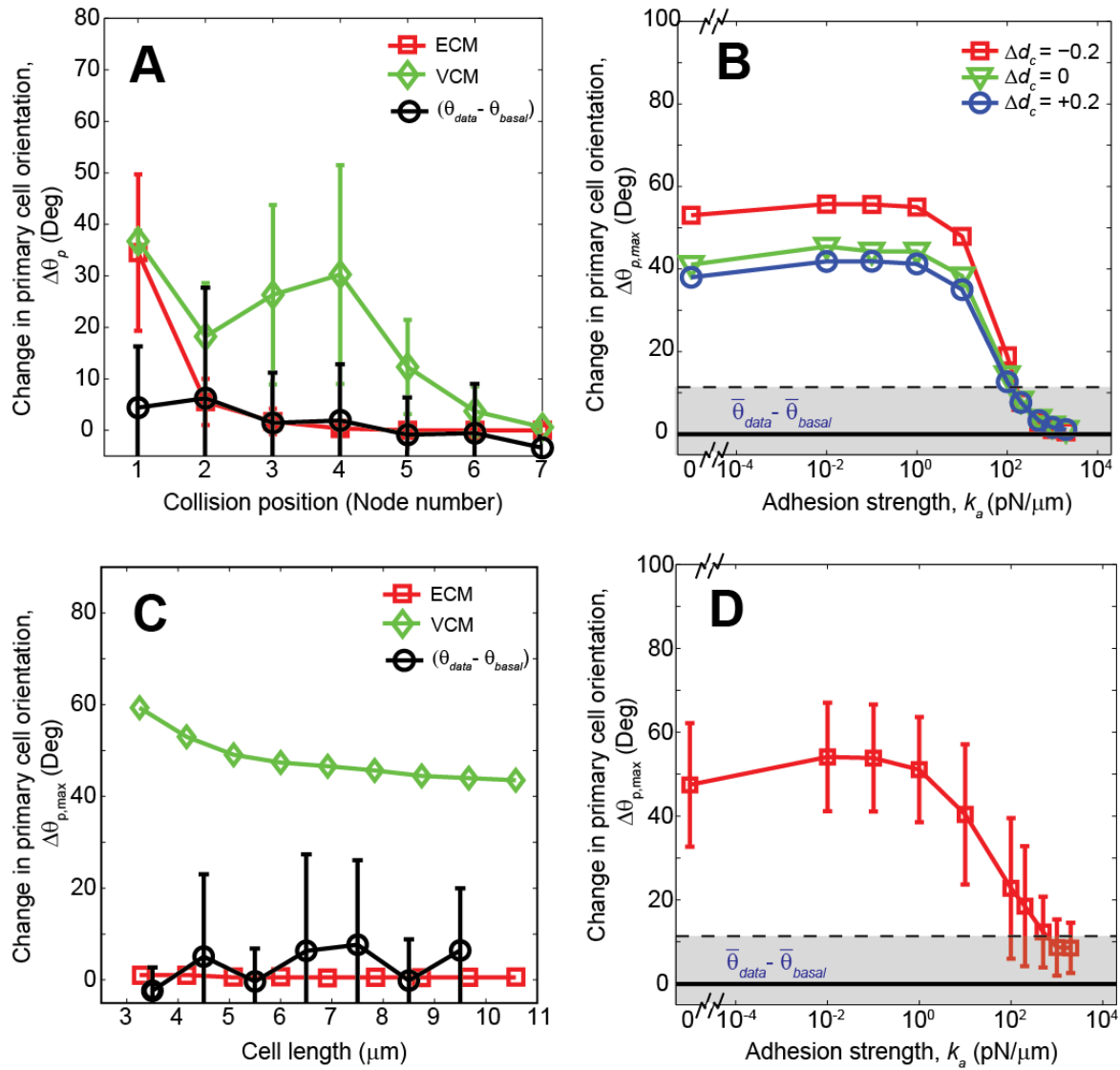

Figure S5: **Cell collision behavior for variations in cell geometrical parameters**. (A) Mean and standard deviations in the primary cell orientation changes as a function of cell collision position for variations in the cell mechanical parameters (see Table S2). Black circles represent mean values from the experimental observations. (B) Variations in the maximum change in the primary cell orientation for small perturbation ($\Delta d_c = \pm 0.2 \; \mu m$, ~ half-cell width) of collision position from center of the node. (C) Maximum change in the primary cell orientation in cell collisions as a function of the variation in cell length. Here length of the primary cell is varied while the secondary cell length is fixed at $7 \; \mu m$. We also find that the results are similar for variation of secondary cell length (data not shown). (D) Mean and standard deviation in primary cell orientation change as a function of adhesive strength for an increased number of adhesive complexes per cell (9 complexes).

**Figure S6**

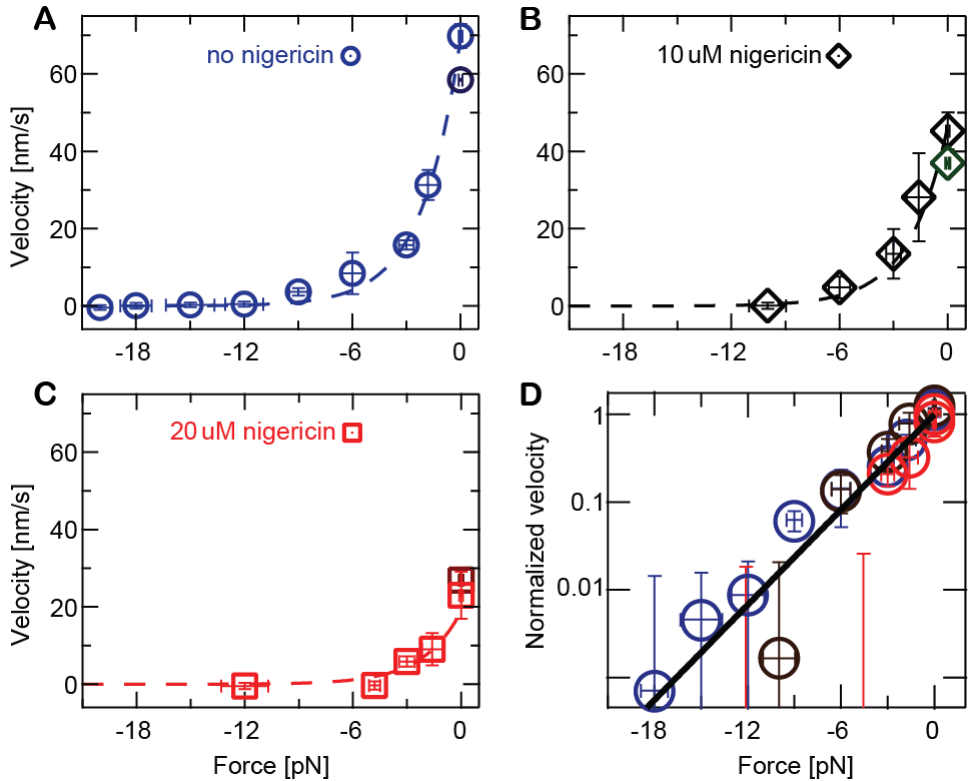

Figure S6: **Force-velocity relation of *M. xanthus* gliding motors at various nigericin concentrations**.(A-C) Force velocity curves for three different nigericin concentrations: 0 µM (A, blue circles), 10 µM (B, black diamonds), 20 µM (C, red squares). Velocity decreases exponentially with force but never becomes negative consistent with an elastic coupling and inconsistent with a viscous coupling between the bead and motor. The dashed lines are exponential fits to the data. Error bars represent the standard error of the mean across trials (> 6 trials per data point). (D) Force-velocity relation on a semi-logarithmic scale with velocities for each nigericin concentration normalized by the load-free velocity. In this representation, the curves collapse onto a single curve with an exponential characteristic force of 2.3 pN.

**Movies SM1, SM2, SM3**

Movie SM1: Video of the cell-cell collision behavior from the VCM model corresponding to the time lapse images shown in Figure 3A.

Movie SM2: Video of the cell-cell collision behavior from the ECM model corresponding to the time lapse images shown in Figure 3B.

Movie SM3: Video of the cell-cell collision behavior from experiments corresponding to the time lapse images shown in Figure 3C.